\documentclass[%
 aip,
rsi,%
 amsmath,amssymb,
 reprint,%
]{revtex4-1}
\usepackage{epsfig}
\usepackage{graphicx}% Include figure files
\usepackage{dcolumn}% Align table columns on decimal point
\usepackage{bm}% bold math
\usepackage[active]{srcltx}
\usepackage{hyperref}
\hypersetup{breaklinks=true}
\newcommand{\be}{\begin{equation}}

\newcommand{\ee}{\end{equation}}
\newcommand{\bea}{\begin{eqnarray}}
\newcommand{\eea}{\end{eqnarray}}

\begin{document}

\Urlmuskip=0mu plus 1mu\relax

\title{Ising Model with conserved magnetization on the Human Connectome: implications on the relation structure-function in wakefulness and anesthesia}
\date{\today}

\author{S. Stramaglia}\affiliation{Dipartimento di Fisica,
Universit\`a degli Studi di Bari, Italy\\}\affiliation{Istituto Nazionale di Fisica Nucleare, Sezione di Bari, Italy\\}\affiliation{Center of Innovative Technologies for
Signal Detection and Processing TIRES,
Università di Bari, Italy\\}
\affiliation{BCAM - The Basque Center for Applied
Mathematics, Bilbao, Spain}
\author{M. Pellicoro}\affiliation{Dipartimento di Fisica,
Universit\`a degli Studi di Bari, Italy\\}
\author{L. Angelini}\affiliation{Dipartimento di Fisica,
Universit\`a degli Studi di Bari, Italy\\}\affiliation{Istituto Nazionale di Fisica Nucleare, Sezione di Bari, Italy\\}\affiliation{Center of Innovative Technologies for
Signal Detection and Processing TIRES,
Università di Bari, Italy\\}
\author{E. Amico}\affiliation{Coma Science Group, University of Li{\`e}ge, Belgium}
\affiliation{Faculty of Psychology and
Educational Sciences, Department of Data Analysis, Ghent University, Belgium\\}
\author{H. Aerts}\affiliation{Faculty of Psychology and
Educational Sciences, Department of Data Analysis, Ghent University, Belgium\\}
\author{J.M. Cort\'es}\affiliation{Ikerbasque, The
Basque Foundation for Science, E-48011, Bilbao, Spain.}
\affiliation{Biocruces Health Research Institute. Hospital
Universitario de Cruces. E-48903, Barakaldo, Spain.\\}
\author{S. Laureys}\affiliation{Coma Science Group, University of Li{\`e}ge, Belgium}
\author{D. Marinazzo}\affiliation{Faculty of Psychology and
Educational Sciences, Department of Data Analysis, Ghent University, Belgium\\}

\date{\today}% It is always \today, today,

\begin{abstract}
Dynamical models implemented on the large scale architecture of the
human brain may shed light on how function arises from the
underlying structure. This is the case notably for simple abstract
models, such as the Ising model. We compare the spin correlations of
the Ising model and the empirical functional brain correlations,
both at the single link level and at the modular level, and show
that their match increases at the modular level in anesthesia, in line with recent results and theories. Moreover, we show that at the peak of the specific heat (the {\it
critical state}) the spin correlations are minimally shaped by the
underlying structural network, explaining how the best match between structure and function is obtained at the onset of criticality, as previously observed.
These findings confirm that brain dynamics under anesthesia shows a departure from criticality and could open the way to novel perspectives when the conserved magnetization is interpreted in terms of an homeostatic
principle imposed to neural activity.

\end{abstract}

%Use showkeys class option if keyword

\maketitle

\textbf{It has been shown that a wide class of models, spanning a wide range from abstract to biologically detailed, reproduce large scale collective dynamics in the brain when they are in a \textit{critical} regime. 
% This phenomenon, whose main signature is the emergence of large scale correlations, arises as a consequence of some conservation law.  
Here we focus on possibly the simplest one, the Ising model, implemented on the structural architecture of the brain, and look at what happens when we introduce a further conservation constraint: the total magnetization remains constant at each step. We show that this leads to an improved correspondence between structure and function at the level of modules. This phenomenon is increased in particular under loss of consciousness, when brain dynamics moves away from the critical regime, thus providing insights on how structure and function interact in the brain.}

\section{Introduction}
One of the key challenges in the study of complex networks is
understanding the relation between structure and the collective
dynamics stemming from it. This issue is of special relevance in
neuroscience, where the question translates to how structurally distinct and distant brain areas dynamically interact
\cite{Park2013}, both in healthy and pathological conditions. Recent
advances in diffusion imaging and tractography methods allow the
noninvasive \textit{in vivo} mapping of white matter cortico-cortical projections at relatively
high spatial resolution \cite{Sporns2011}, yielding a connection matrix
of interregional \textit{structural connectivity} (SC). Similarly, functional MRI can be used to obtain a functional connectivity (FC) matrix, by calculating the statistical dependencies between BOLD time series measured at different sites of the brain. \cite{Shen2015}.  Since the early days of \textit{connectomics}, the relation between SC and FC has been a matter of interest, being expected but not trivial \cite{VandenHeuvel2009a,Hermundstad2013b}.

The intricate interplay between structure and function can be investigated by simulating
spontaneous brain activity on structural connectivity maps. Recent studies \cite{Honey2009,Cabral2011a,Deco2011d,Deco2012d,Deco2012e} have implemented models of dynamical oscillators on the connectome structure \cite{Hagmann2008}. These computational models vary from complex, biologically realistic spiking attractor models, describing the firing rate of populations of single neurons, over mean-field models of neuronal dynamics, down to the simple, biologically-na\"{i}ve Ising model. All these studies agree that  the best agreement of simulated functional connectivity with empirically measured functional connectivity can be retrieved when the brain network operates at the edge of dynamical instability. This state corresponds to the \textit{critical} regime, and for the Ising model coincides with the maximum value of the heat capacity and of the susceptibility. In particular some studies showed that the resting
activity exhibits peculiar scaling properties, resembling the
dynamics near the critical point of a second order phase transition,
consistent with evidence showing that the brain at rest is near a
critical point \cite{Haimovici2013}. Moreover, the possible origin and role
of criticality in living adaptive and evolutionary systems has
recently been ascribed to adaptive and evolutionary functional
advantages \cite{Hidalgo2014}. In \cite{Fraiman2009} the large-scale pattern of empirical
brain correlations was compared with those from a large
two-dimensional Ising model, showing that the match is optimal when
the statistical system is close to the critical temperature. Remarkably, it has been recently argued that propofol-induced sedation and loss of consciousness move brain dynamics away from the critical regime \cite{Tagliazucchi2016}.

However, the Ising model on brain networks has so far been implemented only according to a spin dynamics in which magnetization is not preserved \cite{Glauber1963}. Another class of dynamics exists, in which the total magnetization is preserved: it is used to describe for example alloy systems, where the two different spin states naturally correspond to the two
component atoms that comprise the alloy \cite{Krapivsky2010} and can be implemented via a pair exchange update rule \cite{Kawasaki1966}. If we consider the Ising model on the human connectome as a model of neural activity, the conservation of magnetization may be seen as a sort of homeostatic principle for the overall activity of the brain.

The question we address here is whether an Ising model with conserved magnetization on the human connectome would be a more suitable model for the functional connectivity of brain, in particular under anesthesia, where previous works have hypothesized a departure from criticality. 

Under anesthesia, the brain spans a dynamical repertoire that is reduced with respect to wakefulness. This would result in an increased correspondence between structural and functional connectivity \cite{Barttfeld2015,Tagliazucchi2015a}. Following the reasoning and the results of \cite{Diez2015,Misic2016} we think that this correspondence is to be sought at the level of modules rather than at the level of individual links.

\section{Methods and data}

\subsection{Dynamical Ising models on brain networks}
The study of the Ising model, an appealing description of phase
transitions in ferromagnets, played a fundamental role in the
development of the modern theory of critical phenomena (see, e.g.,
\cite{Huang1987} and the recent review for Ising's model $90^{th}$ birthday \cite{Taroni2015}). In the original model, a regular lattice is populated
by 2-state spins, assuming one of the two values $\sigma_i =\pm 1$.
Pairs of nearest neighbours spins interact so as to favour their
alignment. The Hamiltonian of the system is given by
$${\bf H} = -J \sum_{\langle i j \rangle} \sigma_i \sigma_j, $$ the sum being over nearest
neighbours pairs on the lattice, the positive coupling $J$ favouring
ferromagnetic order. For spatial dimension $d>1$ the model exhibits,
in the thermodynamic limit, a phase transition with finite critical
temperature $T_c$, such that above $T_c$ the spatial arrangement of
spins is disordered with an equal number of up and down spins. Below
$T_c$ the magnetization is non-zero and distant pairs of spins are
strongly correlated. All the equilibrium properties of the Ising
model can be obtained from the partition function $Z=\sum
exp{(-\beta {\bf H}})$, the sum being over the configurations of the
system and $\beta$ being the inverse temperature. Dynamical rules
leading to the same equilibrium are not unique, all the
possibilities being fixed by the detailed balance condition.
Fundamentally, the spin dynamics may or may not conserve the total
magnetization, depending on whether the Ising model is being used to
describe alloy systems, where the magnetization is conserved as it
is related to the composition of the material, or spin systems where
magnetization is not conserved.

According to Glauber dynamics \cite{Glauber1963} the magnetization is not
conserved and each spin is sequentially considered and flipped with
probability $P_{flip}=\left( 1+exp(\beta \Delta E)\right)^{-1}$,
where $\Delta E$ is the energy difference associated to the spin
flip. Here we consider the Kawasaki spin-exchange dynamics
\cite{Kawasaki1966} which conserves the magnetization:  two spins randomly
chosen are swapped with probability $$\exp{(-\Delta E)},$$ where
$\Delta E$ is the variation of the energy corresponding to
exchanging the two spins. A full iteration consists in tentatively
updating all the spins (pairs of spins) for Glauber (Kawasaki)
dynamics.

Turning now to couplings, let us denote $A_{ij}$ the symmetrical
structural connectivity matrix. The Hamiltonian of the Ising model on the network
is $${\bf H} = - \sum_{i,j} J_{ij}\sigma_i \sigma_j,$$ where the
couplings are given by $J_{ij}=\beta A_{ij}$ and the parameter
$\beta$ plays the role of an inverse temperature. Since we deal with
finite size systems, they exhibit a (pseudo)-transition between the
disordered phase and the ordered one, corresponding to the peak of
the specific heat (and of the susceptibility, for the case of
Glauber dynamics).

In \cite{Marinazzo2014} the Ising model with Glauber dynamics
was implemented on the human connectome matrix of \cite{Hagmann2008} at two different spatial scales, 998 and 66 nodes, and the directed and undirected information transfer between nodes was then quantified.

Spin correlations were evaluated using the classical linear Pearson
correlation. The pairwise transfer entropy $TE$, measuring the
information flow from spin $i$ to spin $j$ in each pair connected by
a link in the underlying network was computed as follows:

\begin{multline*}\label{te}
TE_{ij}= \sum_{\sigma_j =\pm 1} \sum_{\Sigma_j =\pm 1} \sum_{\Sigma_i
=\pm 1}  p\left( \sigma_j,\Sigma_j,\Sigma_i \right) \cdot ... \\ 
... \cdot \log{p\left(\sigma_j ,\Sigma_j \right)\; p\left(\Sigma_j,\Sigma_i
\right) \over p\left(\sigma_j,\Sigma_j,\Sigma_i
\right)\;p\left(\Sigma_j \right)},
\end{multline*}
where $p\left(\Sigma_j ,\Sigma_i \right)$ is the fraction of times
that the configuration $(\Sigma_j ,\Sigma_i )$ is observed in the
data set, and similar definitions hold for the other probabilities.

It was shown that at criticality the model displays the maximal
amount of total information transfer among variables, with patterns consistent for both the coarser and the denser scale. Given the fact
that in this line of research we are particularly interested in
information flow in networks, we have verified that the peak of the information transfer corresponds to the peak of the specific heat; therefore a range of temperatures around the peak of the information flow (or, equivalently, the peak of the specific heat) will be taken as the critical regime of the system. 
It is worth mentioning that it has been observed in the regular 2D lattice Ising model that, differently from the pairwise information flow which peaks at criticality, the global information transfer peaks on the disordered side of the transition, and the asymmetry observed in \cite{Barnett2013} as also been observed in \cite{Botcharova2014} when considering both Ising and Kuramoto (from the viewpoint of correlations in phase synchronisation).

\subsection{Data}
The fMRI data that we consider in this work were recorded from
healthy subjects in awake conditions and during propofol anesthesia.
The motivation for the study, the underlying physiological issues,
and the protocol are extensively described in \cite{Boveroux2010}. The
functional MRI (fMRI) data was preprocessed with FSL (FMRIB Software
Library v5.0). The first 10 volumes were discarded for correction of
the magnetic saturation effect. The remaining volumes were corrected for motion, after which slice timing correction was applied to correct
for temporal alignment. All voxels were spatially smoothed with a
6mm FWHM isotropic Gaussian kernel and after intensity
normalization, a band pass filter was applied between 0.01 and 0.08
Hz. In addition, linear and quadratic trends were removed. We next
regressed out the motion time courses, the average CSF signal and the
average white matter signal. Global signal regression was not performed. Data were
transformed to the MNI152 template, such that a given voxel had a
volume of 3mm x 3 mm x 3mm. Finally we obtained 116 time series,
each corresponding to an anatomical region of interest (ROI), by
averaging the voxel signals according to an anatomical template
\cite{Tzourio-Mazoyer2002}. We selected this partition for being the most used
in fMRI connectivity analysis, and because it includes subcortical
structures.
For the diffusion MRI data, we used the publicly available data
contained in the Nathan Kline Institute- Rockland sample described
and downloadable at
\url{http://fcon\_1000.projects.nitrc.org/indi/pro/eNKI\_RS\_TRT/FrontPage.html}.
As a first step, the images were corrected for motion and eddy currents due to changes in the gradient field directions of the MR scanner. In particular, the eddy-correct tool from FSL was used to correct both artifacts, using an affine
registration to a reference volume. After this, DTIFIT was used to
perform the fitting of the diffusion tensor model for each voxel. Then, a deterministic tractography algorithm
(FACT) \cite{Mori1999} was applied using TrackVis \cite{Wang2007}, an
interactive software for fiber tracking. Two computations were performed to
transform the anatomical atlas to each individual space: (1) the transformation
from the MNI template to the subject's structural image (T1), and
(2) the transformation from the subject's T1 to the subject's diffusion image space.
Combining both transformations, each atlas region is transformed to
the subject's diffusion space, allowing to count the number of reconstructed streamlines
connecting all ROI pairs.

It is worth to note that for group level analyses at the scales
considered in this study it is not relevant that the structural
connectivity used for the simulations is not the one obtained from
the same subjects for which the functional connectivity was computed and compared \cite{SanzLeon2013}.

\subsection{Modularity and modular similarity between networks}

A key concept in network theory is modularity \cite{Newman2006}. It
describes how efficiently a network can be partitioned in
sub-networks, and it is particularly relevant when it
comes to study the interplay between anatomical segregation and
functional integration in the brain \cite{Sporns2015a}. Maximization
of the modularity Q allows to identify communities. Here we use the
algorithm described in \cite{Rubinov2011} which also takes into account negative weights, a situation that frequently arises in functional networks. The
resolution parameter is set to its default value $\gamma =1$
and, for each network, the algorithm was run 1000 times choosing the
maximal Q and the corresponding partition. In order to compare, at
the modular level, two networks with the same nodes, we calculate
the similarity of their partitions quantifying it by the mutual
information approach described in \cite{Meila2007}. The code
used to compute these quantities is contained in the Brain Connectivity Toolbox
(\url{https://sites.google.com/site/bctnet/})

\section{Results}
\begin{figure}[ht!]
\begin{center}
\epsfig{file=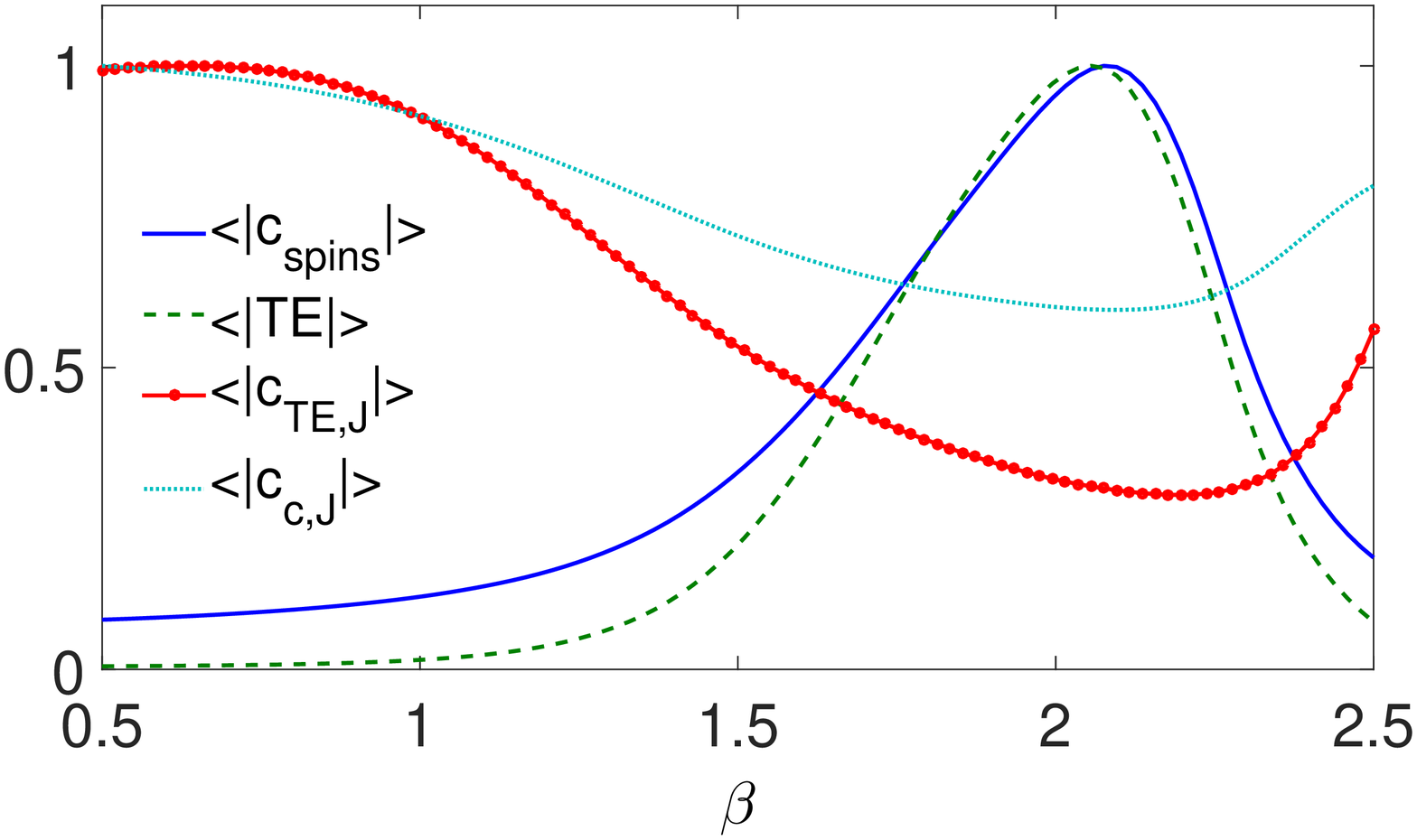,width=9.cm}
\end{center}
\caption{{\small The Ising model with Kawasaki dynamics is
implemented on the 116 regions structural connectome. The following
quantities are depicted as a function of the inverse temperature $\beta$: the correlation
between the value of TE and J over all the anatomically connected
pairs (red line, with bullets); the correlation between the value of c and J over
all the anatomically connected pairs (cyan, dotted line); the normalized average
absolute value of the correlation (blue, full line); the normalized average transfer
entropy (green, dashed line).\label{fig1}}}
\end{figure}

We implemented the Ising model on the structural brain
networks, and evaluated the transfer entropy and the spin correlations, as described in the previous section.
The results shown here correspond to the average over 1000 runs, each consisting of 30000 full iterations of the lattice (after discarding the transient). 
Since we deal with a small system, and we are not interested in the low temperature limit, we always assumed zero magnetization for Kawasaki dynamics, i.e. the starting configuration consisted in an equal number of plus and minus spins randomly assigned to nodes. In other words, due to the small size, we assume vanishing equilibrium global magnetization.
We start considering, in figure (\ref{fig1}), the following problem:
to which extent are the functional patterns of the Ising systems
shaped by the underlying topology, at the level of individual links?
We compare the anatomical network with the functional networks
provided by the dynamical system, as a function of the inverse
temperature. The following quantities are depicted, derived from the Ising
model with Kawasaki dynamics on the structural connectome
corresponding to awake conditions: the Pearson correlation between
the transfer entropy TE and the coupling \textit{J} over all the anatomically
connected pairs, and the Pearson correlation between spin
correlations \textit{c} and \textit{J} over all the anatomically connected pairs. The
plots of the average absolute value of the correlation and the
average transfer entropy, both peaking at criticality, are also
displayed for comparison. As figure (\ref{fig1}) shows, in the temperature regime at which information transfer and correlations are higher (identified with the critical regime, see below), the link correlation of TE and \textit{c}, with \textit{J}, is minimal;
in other words the critical states appear to be the ones at which
the functional pattern is minimally shaped by the details of the
underlying structural network. These findings are in line with a previous study on Ising models implemented on the connectome, in which the best fit between model and empirical correlations was observed when the entropy of the model attractors started to increase, and not at its peak \cite{Deco2012e}.

Next, in order to elucidate whether and how the fit between model and empirical correlation change with the level of consciousness, we consider both the
fMRI data recorded from healthy subjects in awake conditions and
during propofol anesthesia, as well as the corresponding structural data.
Varying the temperature, we implemented the Ising model with Kawasaki
dynamics on the structural architecture connecting the 116 ROIs. We compare the
corresponding spin correlations with the empirical functional
correlations.

\begin{figure}[ht!]
\begin{center}
\epsfig{file=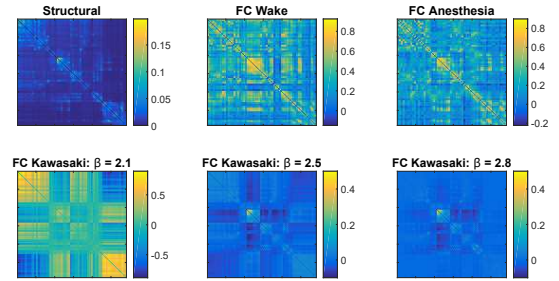,width=9.cm}
\end{center}
\caption{{\small Top: matrices corresponding to empirical data. Structural connections (left), average correlations in wakefulness (center), average correlations in anesthesia (right).\\Bottom: average correlations in simulated time series for three values of the inverse temperatures, corresponding to relevant points in the curves of the following figures.\label{fig2}}}
\end{figure}

In order to visualize the typical patterns of the connectomes under exam, in figure (\ref{fig2}) we depict
the 116 $\times$ 116 structural connectivity matrix, the empirical
functional connectivity matrix for wake and anesthesia conditions, and the
patterns of correlations of the Kawasaki model tuned at three different temperatures corresponding to relevant regimes in the curves described below (greatest linkwise correlation, greatest mean squared error, maximum mutual information between structural and functional modules).

\begin{figure}[ht!]
\begin{center}
\epsfig{file=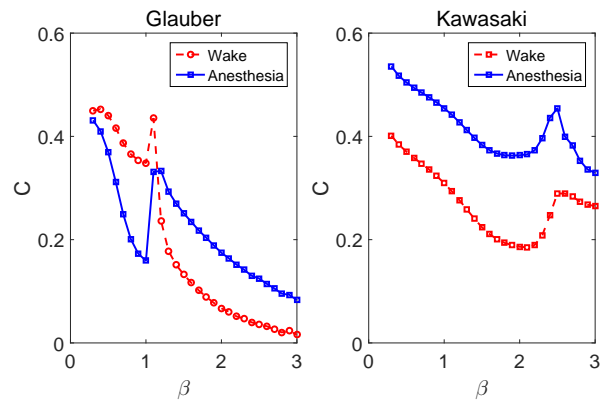,width=9.cm}
\end{center}
\caption{{\small  The link-wise correlation between the model spin
correlations and the empirical functional connectivities is depicted
as a function of the inverse temperature $\beta$ for wakefulness
(dashed red line) and anesthesia (full blue line). Left panel:
Glauber dynamics. Right panel: Kawasaki dynamics. \label{fig3}}}
\end{figure}

As depicted in figure (\ref{fig3}), when the link-wise correlation
between model and empirical functional patterns is considered, the
match between model and empirical correlations is higher in wakefulness for Glauber dynamics, and in anesthesia for Kawasaki dynamics, in the respective critical regimes for each dynamics. Moreover, under sedation, the Kawasaki dynamics results in a better fit and in a clearer separation between wake and anesthesia, compared to the Glauber dynamics.

\begin{figure}[ht!]
\begin{center}
\epsfig{file=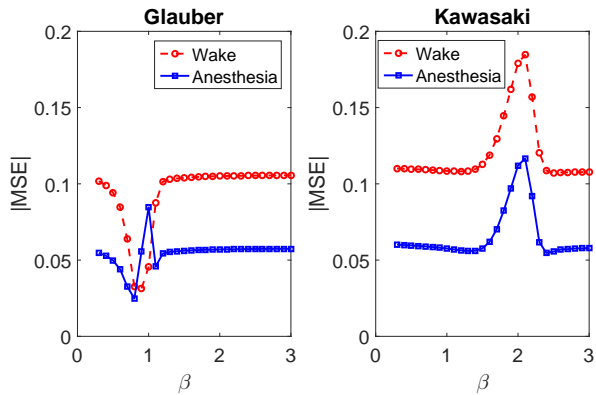,width=9.cm}
\end{center}
\caption{{\small The mean square error between the model spin
correlations and the empirical functional connectivities is depicted
as a function of the inverse temperature $\beta$ for wakefulness
(dashed red line) and anesthesia (full blue line). Left panel:
Glauber dynamics. Right panel: Kawasaki dynamics.  \label{fig4}}}
\end{figure}

In figure (\ref{fig4}), the match between
empirical and spin correlations is measured in terms of the mean
square error between the two patterns, i.e. the average of
$\left(c_{ij}^{spin} -c_{ij}^{empirical}\right)^2$ over all
 pairs of brain regions, $c_{ij}^{spin}$ being the spin
 correlation of the Ising models and $c_{ij}^{empirical}$ the
 empirical functional connectivity. Results show that again, the best match  for anesthesia is better than that for wake conditions, both for Glauber and Kawasaki
 dynamics. Using the mean square error, however, Glauber dynamics showed a  better fit.

\begin{figure}[ht!]
\begin{center}
\epsfig{file=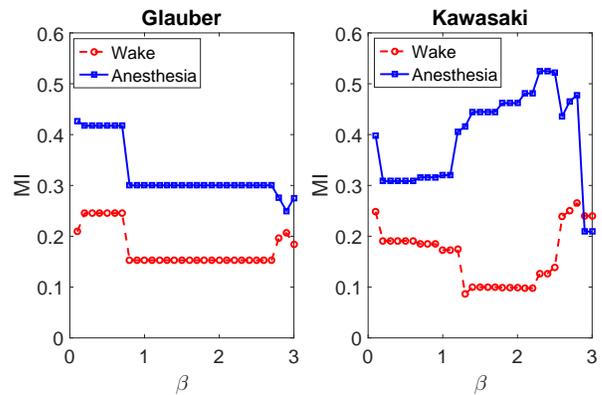,width=9.cm}
\end{center}
\caption{{\small The mutual information between the partitions of
the empirical and model functional networks, as a function of the
inverse temperature $\beta$ is depicted for Kawasaki and Glauber
dynamics, and for wake and anesthesia conditions. The modular
decomposition is obtained by maximization of modularity.
\label{fig5}}}
\end{figure}

However, it has been shown that a modular comparison is
better suited to investigate the interplay between structure and function
\cite{Diez2015}. Hence, in order to better describe the relation
between empirical and simulated functional patterns at the
modular level, we have evaluated the mutual information between
partitions (obtained by maximizing the modularity)  as a function of
$\beta$ to quantify the relation between empirical functional
correlations and spin correlations from the model,  for wake
subjects and subjects under anesthesia. In figure (\ref{fig5}) the
mutual information is plotted for both Kawasaki and Glauber
dynamics. In wakefulness, where the same structural connectivity subserves a wide repertoire of activity, we observe a reduction of the mutual information in the critical regime with respect to the disordered phase, both for Glauber and Kawasaki dynamics. This behavior is reversed, leading to increased mutual information between structural and functional models, for the Kawasaki dynamics under sedation.

These results speak to the fact that Ising models tuned at criticality result in connectivity matrices with a generally good linkwise resemblance with the empirical ones, comparable or even better than the one obtained with more biologically precise models \cite{Deco2013,Falcon2015}.
This similarity is represented by a peak in the curves in figures \ref{fig3} which follows a trough, corresponding to the peak in mean squared error of \ref{fig4}.
According to the pairwise metric though, the best resemblance remains the not-so-interesting one corresponding to the limit of high temperatures.
On the other hand, if we look at the mutual information between structural and functional modules, we can observe that, with respect to the disordered phase, Kawasaki dynamics in the critical regime leads to a decreased match between structural and functional modules in wakefulness, and an increased one in anesthesia. These results, in line with our hypotheses, are not evidenced by Glauber dynamics.

\section{Discussion and conclusions}

We have considered pair exchange update rules for the Ising model
\cite{Kawasaki1966} implemented on the structural brain network at
the macroscale, using a data set of healthy subjects scanned during
quiet wakefulness and during deep sedation, a condition in which the
structure-function relation is modified. Our results show that the
structure-function relation is strengthened under anesthesia, both
at the link and modular level, compared to wake conditions. Having shown that at criticality the functional pattern is less dependent on the underlying structural network, it follows that anesthesia takes the brain dynamics farther from the critical regime, in accordance with previous evidence.

Moreover, at the modular level we obtained a better match with
empirical functional correlations using Kawasaki dynamics compared to the more common Glauber dynamics for which the magnetization is not preserved
\cite{Glauber1963}. This improved match suggests that the
conservation law of the Kawasaki dynamics might admit a
physiological counterpart. A possible interpretation is seeing it as
an {\it effective} implementation, due to time scales separation, of
the coupling from metabolic resources to neural activity, a key
ingredient that is missing in neural models on the connectome
\cite{Roberts2014}.

In agreement with recent theoretical frameworks \cite{Moretti2013},
our results suggest that a wide range of temperatures correspond to
{\it criticality} of the dynamical Ising system on the connectome,
rather than a narrow interval centered in a critical state. In such
critical conditions, the correlational pattern is minimally shaped by
the underlying structural network. It follows that, assuming that the
human brain operates close to a {\it critical} regime
\cite{Chialvo2010}, there is an intrinsic limitation in the
relationship between structure and function that can be observed in
data. We have shown that empirical correlations among brain areas
are better reproduced at the modular level using a model which
conserves the global magnetization. The most suitable way to compare
functional and structural patterns is to contrast them at the
network level, using, e.g., the mutual information between
partitions like in the present work.

During the awake resting state, spontaneous brain activity
constantly fluctuates across brain regions, exhibiting a rich
repertoire of functional connectivity patterns. Previous studies
accounted for long-range resting-state functional connectivity
persisting even after loss of consciousness
\cite{Vincent2007,Fernandez-Espejo2012,Guldenmund}. A
recent study on monkeys \cite{Barttfeld2015} posited that the role of
structural connectivity in sculpting functional connectivity maps
changes during wakefulness and anesthesia. According to the authors,
wakefulness seems to be characterized by a rich repertoire of
connectivity patterns, while the functional connectivity patterns
under sedation follow the underlying brain structure. Another study reports increased similarity between whole-brain anatomical and functional connectivity networks during deep sleep \cite{Tagliazucchi2015a}. Our results,
showing that the structure-function correspondence is enhanced under
anesthesia at the modular level, are in accordance with this evidence.

Summarizing, we have considered the Ising model as a valuable tool
to explore large scale brain dynamics. We have shown that the conservation of
magnetization leads to better correspondence between structure
and function, notably where this is more expected, that is in loss of consciousness, again speaking to a less critical dynamical regime. 
The reason of this improvement might lie in the features added by the Kawasaki dynamics to large scale connectivity (i.e. negative correlations). Its biological counterparts could be intuitively found in some homeostasis mechanism or metabolic constraint, but no validation tools for this conjecture exist at the moment. 

Finally, our results confirm that matches between structure and function should be sought at the modular level rather than among individual links, in line with recent work \cite{Diez2015,Misic2016,Kujala2016}.
%
%\begin{itemize}
%\item The Ising model is a valuable tool to explore large scale brain dynamics
%\item Conservation of magnetization in the Ising model adds feature to large scale connectivity (i.e. negative correlations). Its biological counterparts could be intuitively found in some homeostasis mechanism or metabolic constraint, but no validation tools for this conjecture exist at the moment.
%\item Conservation of magnetization in the Ising model leads to better correspondence between structure and function, notably where this is more expected, that is in loss of consciousness
%\item This correspondence has to be sought at the modular level rather than among individual links.
%\end{itemize}
%

%
%\begin{figure}[ht!]
%\begin{center}
%%\epsfig{file=figure6_cmreal.eps,height=6.cm}
%\end{center}
%\caption{{\small (a) Cross-modularity between structural connectome
%and averaged empirical functional connectivity for wakefulness and
%anesthesia as a function of the number of modules. (b)
%Cross-modularity between structural connectome and simulated
%functional connectivity as a function of the number of modules and
%of the inverse temperature $\beta$. \label{fig6}}}
%\end{figure}
%

\bibliography{isingconn}

\end{document}